\documentclass[rsi,reprint]{revtex4-2}

\usepackage{amsmath}
\usepackage{amssymb}

\usepackage{natbib}

\setcitestyle{numbers}

\usepackage{graphicx}
\usepackage{placeins} 
\usepackage{dcolumn}
\usepackage{bm}

\usepackage[utf8]{inputenc}
\usepackage[T1]{fontenc}
\usepackage{mathptmx}

\usepackage{hyperref}
\hypersetup{colorlinks=true, pdfstartview=FitV, linkcolor=blue, citecolor=blue, plainpages=false, pdfpagelabels=true, urlcolor=blue}

\usepackage[all]{hypcap}

\begin{document}



\title{Total-Reflection High-Energy Positron Diffractometer at NEPOMUC \textendash{}\\Instrumentation, Simulation and First Measurements} 


\author{Matthias Dodenhöft$^\ast $, Sebastian Vohburger, and Christoph Hugenschmidt}
\affiliation{Technische Universit\"at M\"unchen and Forschungs-Neutronenquelle Heinz Maier-Leibnitz FRM II, Lichtenbergstr. 1, 85748 Garching, Germany}


\date{\today}

\begin{abstract}
$\newline$
We report the instrumentation of a new positron diffractometer that is connected to the high-intensity positron source NEPOMUC. Crucial elements for the adaption of the positron beam are presented, which include the magnetic field termination, the optional transmission-type remoderator for brightness enhancement and the electrostatic system for acceleration and beam optics. The positron trajectories of the remoderated and the twofold remoderated beam have been simulated to optimize the system, i.e. to obtain a coherent beam of small diameter. Within a first beamtime we tuned the system and characterized the direct beam. For the twofold remoderated beam of 10$\thinspace$keV energy, we experimentally observe a beam diameter of \textit{d} < 1.3$\thinspace$mm, which agrees well with the simulation.
\end{abstract}

\maketitle


\section{Introduction}
\label{sec:Mot}


The properties of a solid surface are determined by its elementary composition, the chemical bonds and the surface structure \citep{Woo94}. In particular, the crystalline surface structure plays a crucial role for a variety of physical phenomena and is important for many technical processes, such as epitaxial growth of ultra-thin films \citep{Ono03} or heterogeneous catalysis \citep{Hua19}. Moreover, realistic and accurate models of the surface structure are necessary to calculate the electronic band structure at the surface using density functional theory (DFT) \citep{Lut10}. Surface and bulk crystal structure typically differ since the atoms at the surface experience a different crystal potential and form distinct chemical bonds at the interface. This becomes evident by reconstructions, i.e. the rearrangement of surface atoms that changes the periodicity of the crystal structure at the surface. For a certain material, there can be a multitude of different surface structures \citep{Rus08,Kie11} that often involve the interaction with gas molecules \citep{Har06} and depend on external parameters such as temperature or pressure to minimize the free energy \citep{Yeu17}. For silicon and only for the surfaces with low Miller indices, over 300 different structures with and without adsorbates have already been found experimentally \citep{Lif95}. Many reconstructions involve more than just a single atomic layer at the topmost surface, so that also subsurface layers can substantially differ from the bulk structure \citep{Lut10}. In the simplest case, subsurface layers only exhibit lattice relaxations.$\newline$

The precise determination of a surface structure can thus be challenging and the combination of complementary techniques is often required to obtain a comprehensive understanding \citep{Zan88,Vic09}. The local structure can be investigated in real space by scanning probe techniques such as Scanning Tunneling Microscopy (STM) \citep{Bin82,Bin83} or Atomic Force Microscopy (AFM) \citep{Bin86}. These techniques feature ultimate surface sensitivity, but cannot probe sublayer structures and related phenomena. Moreover, chemical and topological information is superimposed, which complicates the interpretation. In contrast, diffraction techniques probe the long-range periodic order in reciprocal space. Surface X-ray diffraction (SXRD) \citep{Eis81a,Rob86,Fei89} is conducted in grazing incidence close to the critical angle of total external reflection, which is usually in the range of 0.2\thinspace-\thinspace 0.6\thinspace$^{\circ}$ \citep{Kov95}. The weak interaction and large penetration depth of X-rays apparently hamper the analysis of the surface structure as the signal is superimposed by a large background from the bulk. As multiple scattering is negligible, SXRD can be well described by kinematic diffraction theory \citep{Taj20}. The most commonly applied technique for surface structure analysis is Low-Energy Electron Diffraction (LEED) \citep{Dav27a,Dav27b}. Besides LEED, Reflection High-Energy Electron Diffraction (RHEED) is widely used, especially in combination with molecular beam epitaxy (MBE) to not only study the surface structure \citep{Shi04}, but also to monitor growth rate \citep{Har81,Nea83}. The surface sensitivity of LEED and RHEED mainly stems from the large cross section for inelastic scattering. Electrons that penetrate deeper into the sample are more likely scattered inelastically, but only coherent, elastically scattered electrons contribute to the Bragg diffraction pattern. Using electron diffraction, it is possible to access both, the periodic order of topmost and subsurface layers. However, the quantitative analysis of the superimposed signal can be difficult, especially when structural models have a lot of free parameters. For LEED and RHEED, dynamical diffraction theory is necessary to calculate the intensities of diffracted beams \citep{Ich04}.

More recently, Total-Reflection High-Energy Positron Dif-fraction (TRHEPD) demonstrated to be an ideal technique to determine the crystalline structure of topmost and subsurface atomic layers \citep{Fuk13a,Fuk18}. However, up to now, there has been just a single TRHEPD experiment in operation worldwide. In order to benefit from the high beam intensity available at \mbox{NEPOMUC} in Munich, we developed a new positron diffractometer.

In this paper, we introduce the fundamental features and advantages of TRHEPD and discuss different aspects of the instrumentation. In particular, we emphasize differences to the operational setup in Japan, which are related to the different properties of the positron sources. Finally, we present the results from a first beamtime that are compared with simulations to characterize the direct beam.

\section{Total-Reflection High-Energy Positron Diffraction}
\label{TRHEPD intro}

\begin{figure}
\includegraphics[trim = 15mm 160mm 25mm 50mm, clip, width=7.5cm]{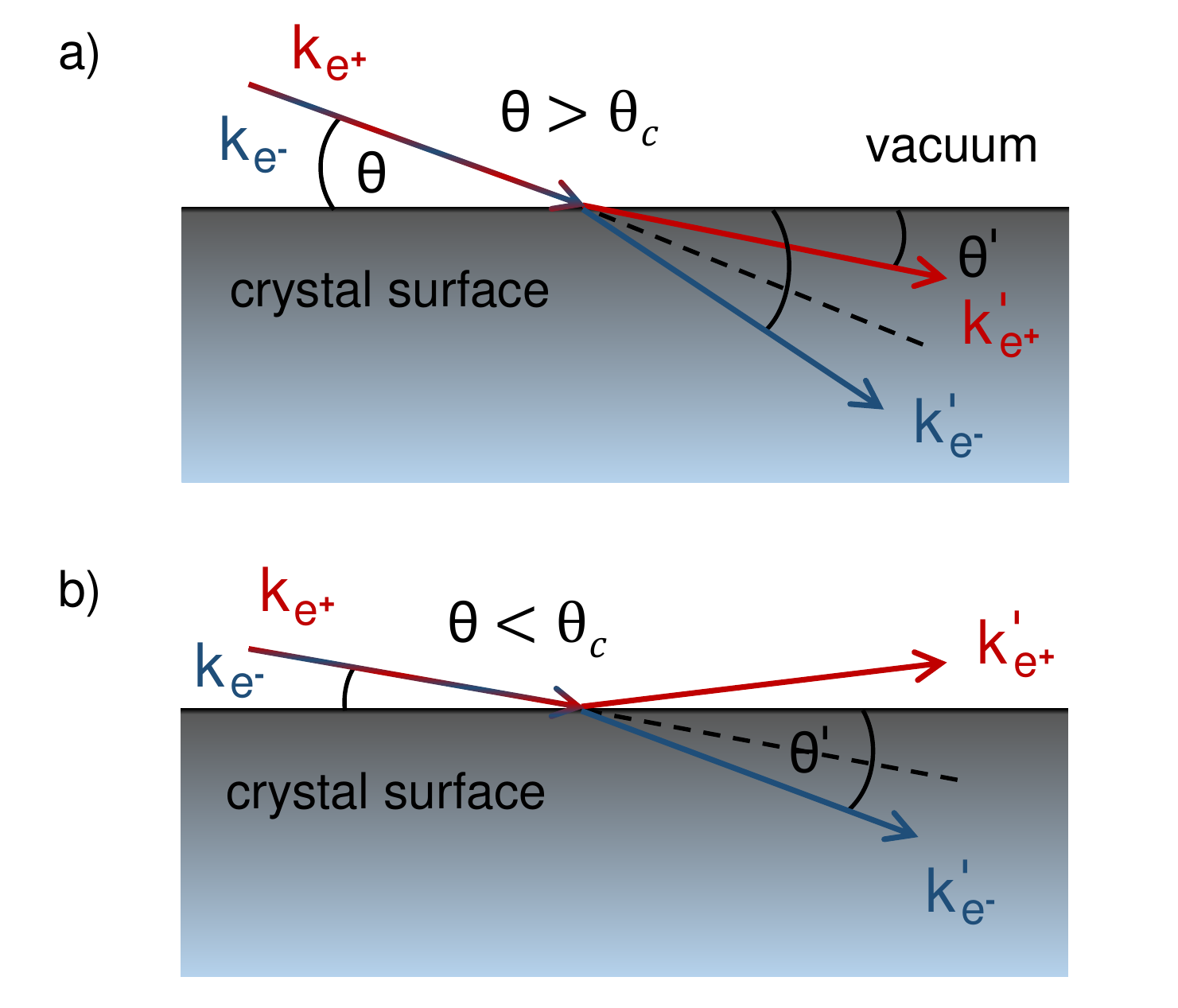}
\setlength\belowcaptionskip{-0cm}
\caption{\label{fig:Electron_Positron_Beugung} Electrons and positrons in grazing incidence (schematic). a) For $\protect\theta$ > $\protect\theta$\textsubscript{c} incoming positrons are refracted towards the crystal surface. b) For $\protect\theta$ < $\protect\theta$\textsubscript{c} positrons experience total reflection. In comparison, electrons are always refracted into the crystal.}
\end{figure}

TRHEPD is the positron counterpart of RHEED and has been suggested by A. Ichimiya in 1992 \citep{Ich92}. As for RHEED, the surface sensitivity of TRHEPD primarily originates from the large cross section for inelastic scattering. However, in contrast to electrons, incident positrons are repelled from the surface as the crystal potential \textit{V} is repulsive \citep{Hug16}. Without considering relativistic effects, the wave vector \textit{k} of the incident beam is
\begin{equation}
k = \frac{\sqrt{2m\textsubscript{e}E\textsubscript{0}}}{\hbar},
\end{equation}
where \textit{m\textsubscript{e}} is the electron or positron mass, \textit{E\textsubscript{0}} is the kinetic energy of the incident beam and $\hbar$ is the reduced Planck constant. Inside the crystal, the wave vector \textit{k$^\prime$} can be expressed as
\begin{equation}
k^\prime = \frac{\sqrt{2m\textsubscript{e}E}}{\hbar}  =  \frac{\sqrt{2m\textsubscript{e}(E\textsubscript{0}-qV)}}{\hbar},
\end{equation}
where \textit{qV} is the average potential energy inside the crystal. To ensure continuity across the interface, the parallel components of the wave vector have to be equal:
\begin{equation}
k^\prime\textsubscript{$\parallel$} = k\textsubscript{$\parallel$}\ \Rightarrow\ k^\prime\cos{\theta^\prime} = k\cos{\theta}.
\end{equation}
Using Snell's law, we can calculate the refractive index \textit{n} from the angle of incidence $\theta$ and the refracted internal glancing angle $\theta^\prime$ \citep{Fuk18}:
\begin{equation}
n = \frac{\cos{\theta}}{\cos{\theta^\prime}} =  \frac{k^\prime}{k} = \sqrt{1-\frac{qV}{E\textsubscript{0}}}.
\end{equation}
For positrons we find $qV=eV>0$ and thus $n < 1$. Consequently, positrons are refracted towards the crystal surface as illustrated in Fig. \ref{fig:Electron_Positron_Beugung}$\thinspace$a). In contrast, $qV=-eV<0$ for electrons, which leads to $n > 1$, i.e. refraction away from the surface. Moreover, unlike electrons, positrons exhibit total reflection for angles of incidence $\theta < \theta\textsubscript{c}$, as depicted in \mbox{Fig. \ref{fig:Electron_Positron_Beugung}$\thinspace$b)}. Total reflection takes place if the kinetic energy of the positron beam perpendicular to the surface is smaller than the potential energy \textit{eV} inside the crystal. The critical angle $\theta\textsubscript{c}$ can thus be calculated as
\begin{equation}
\theta\textsubscript{c} = \arcsin{\sqrt\frac{eV}{E\textsubscript{0}}}.
\label{eq:total_reflection}
\end{equation}
In the regime of total reflection, TRHEPD features outstanding surface sensitivity \citep{Fuk14} and by tuning the angle of incidence above $\theta\textsubscript{c}$, it is also possible to obtain information about subsurface layers. Using equation \eqref{eq:total_reflection} we calculate $\protect\theta$\textsubscript{c} $\approx$ 2$^{\circ}$ for a 10$\thinspace\thinspace$keV positron beam incident on a Si surface  with \mbox{\textit{eV} = 12$\thinspace\thinspace$eV \citep{Hug16}}. This angular range is experimentally well accessible. We emphasize that the critical angle for SXRD is roughly an order of magnitude smaller (see \mbox{section \ref{sec:Mot}}).

As for electrons, dynamical effects cannot be neglected for the calculation of TRHEPD intensities, even not in the case of total reflection \citep{Hay04}. However, due to the reduced penetration depth of positrons, the interaction with the solid is less complex, which typically improves the agreement between theory and experiment \citep{Fuk14}. Pronounced Kikuchi lines that stem from inelastic and subsequent elastic scattering can complicate the analysis of RHEED patterns, but are practically absent for TRHEPD \citep{Hyo14,Mit20}.

In 1998, Kawasuso and Okada reported the first TRHEPD pattern using a \textsuperscript{22}Na-based, moderated and collimated positron beam with an intensity of $\sim\thinspace$5000$\thinspace\thinspace$e\textsuperscript{+}/s \citep{Kaw98}. After an upgrade, this setup has mainly been used to investigate various surface structures induced by metal adatoms on Si(111) \citep{Fuk06b,Fuk07a,Fuk07b,Has09,Fuk12} and Ge(111) \citep{Fuk06a,Fuk08}. Some of these structures could not be characterized using conventional techniques while TRHEPD contributed to the development of novel or refined structural models. In 2010, the experimental setup was moved to the Slow Positron Facility (SPF) at KEK in Japan \citep{Wad12} and later replaced by an improved, brightness-enhanced setup \citep{Mae14}. Both systems are described in detail by Fukaya \emph{et al.} in \citep{Fuk18}. At the SPF, a dedicated 55$\thinspace$MeV linac is used to produce a pulsed positron beam with a kinetic energy of up to 35$\thinspace$keV, an initial intensity of $\sim\thinspace$5$\cdot$10$\textsuperscript{7}\thinspace\thinspace$e\textsuperscript{+}/s and more than $\sim\thinspace$5$\cdot$10$\textsuperscript{5}\thinspace\thinspace$e\textsuperscript{+}/s after transport and remoderation \citep{Fuk14}. The high intensity allows sample orientation in real time and enhances the signal to noise ratio significantly. Using this setup, the surface structure of rutile-TiO$\textsubscript{2}$(110)$\thinspace$-$\thinspace$(1$\times$2) - under debate for over 30 years - could be determined \citep{Moch16}. Recent measurements also comprise the analysis of 2D materials \citep{Fuk18b}, namely the precise determination of the spacing between graphene and different metal substrates \citep{Fuk16a}, the buckling and bonding angles of silicene \citep{Fuk13b} and the analysis of germanene \citep{Fuk16b}. Moreover, for superconducting Ca-intercalated bilayer graphene grown on SiC(0001), it was found that the Ca atoms do not intercalate between the two graphene layers as expected, but between the graphene-buffer interlayer \citep{End20}.

\section{Instrumentation of the new positron Diffractometer at NEPOMUC}
\label{sec:Inst}

Inspired by the Japanese pioneering work, we report the instrumentation of a new TRHEPD experiment. To the best of our knowledge, this is the second setup in the world. It operates at the neutron induced positron source Munich (NEPOMUC) located at the research reactor FRM II. NEPOMUC is the world's most intense positron source and provides a monoenergetic beam of more than 10$\textsuperscript{9}\thinspace\thinspace$e\textsuperscript{+}/s \citep{Hug14}.

\subsection{Experimental Boundary Conditions}
\label{Beam_Properties}

At NEPOMUC, positrons are generated via pair production from high-energy $\gamma$-rays emitted from \textsuperscript{113}Cd after thermal neutron capture \citep{Hug14}. Based on the reactor operation, the positron beam is continuous, in contrast to the pulsed beam available at the SPF. A pulsed structure of the beam is not beneficial for TRHEPD as the intensity of a pulse is far too low for time-resolved measurements. Conversely, a continuous beam is favorable as it doesn't saturate the detector. This enhances the detection efficiency and is a prerequisite for quantitative rocking curve analysis. On these grounds, it is also planned to equip the TRHEPD beamline at the SPF with a pulse stretcher in the near future \citep{Hyo20}.

Besides intensity other beam parameters such as diameter, divergence and energy spread are crucial to obtain a coherent positron beam suitable for diffraction \citep{Kaw04}. The primary NEPOMUC beam has a diameter of $\sim\thinspace$9$\thinspace$mm at full width half maximum (FWHM) and an energy spread of 50$\thinspace$eV perpendicular to the direction of propagation. For this reason, we use the brightness-enhanced, remoderated NEPOMUC beam for diffraction. The remoderator stage consists of a W(100) single crystal in reflection geometry and a movable crystal holder that has recently been upgraded \citep{Dic20}. The remoderated beam has an intensity of $\sim\thinspace$5$\cdot$10$\textsuperscript{7}\thinspace\thinspace$e\textsuperscript{+}/s with a beam diameter of less than 2$\thinspace$mm FWHM. After remoderation, the beam is guided adiabatically to the TRHEPD setup by a longitudinal magnetic field of typically 4$\thinspace$mT. A second remoderator in front of the experiment can be used to increase the brightness of the beam even further (see section \ref{Remoderation}).

To reach the final energy, the positron beam is accelerated by an electrostatic potential. Early TRHEPD measurements in Japan have been conducted using a 20$\thinspace$keV beam \citep{Kaw98}, while the recent setup at the SPF is optimized for a remoderated beam with fixed energy of 10$\thinspace$keV \citep{Mae14}. At the SPF, the positron source is set to a (positive) high voltage of 15$\thinspace$kV for TRHEPD to work in a sample environment that is grounded. Due to the close vicinity to the reactor and the limited space available, the NEPOMUC source is not designed to be set to such high voltages. Consequently, a high negative potential has to be applied at the experimental station being a main challenge for the instrumentation. After acceleration, the final potential must be preserved until the diffracted positrons are detected, i.e. this region has to be field-free. Therefore, the whole beam path is enclosed by a Faraday cage, which also includes the sample stage, the top of the positioning system and the micro channel plate (MCP) assembly for detection. Additionally, all customized solutions have to satisfy the high requirements regarding ultra-high vacuum (UHV) condition to be suitable for surface physics.

\subsection{Overview of the Munich Positron Diffractometer}
\label{Features}

\begin{figure*}
\includegraphics[trim = 5mm 0mm 15mm 0mm, clip, width=18cm]{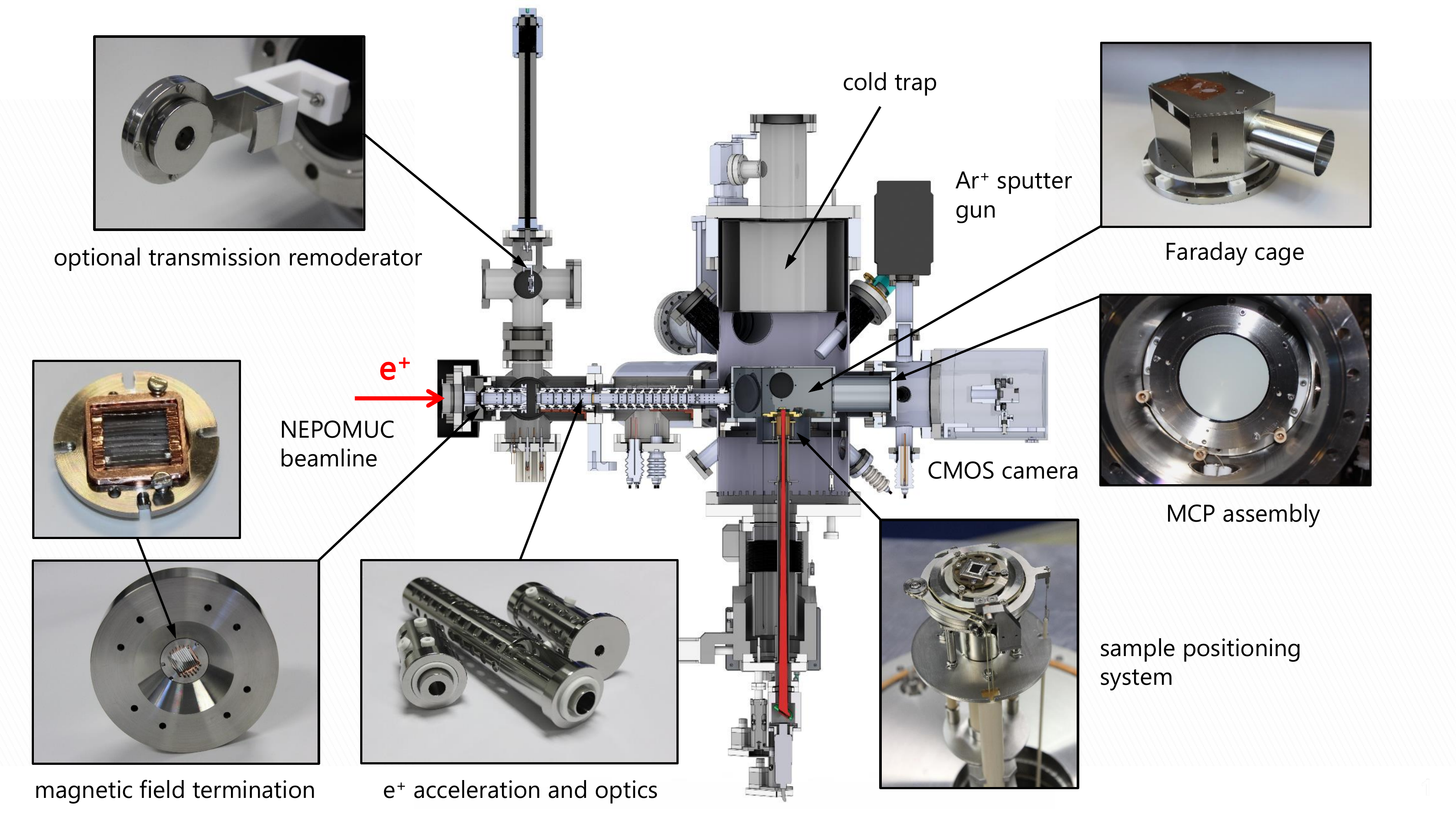}
\caption{\label{fig:TRHEPD_overview} Cross-sectional view of the TRHEPD experiment with photos of the key components. The positron beam enters via the NEPOMUC beamline from the left, by passing a magnetic field termination before being accelerated by a system of electrodes. It is possible to focus the beam onto a Ni(100) foil to increase the brightness via remoderation. After acceleration, the beam enters the field-free region within the Faraday cage where it is diffracted by the sample. The signal is amplified by a MCP assembly and visualized by the attached phosphor screen, so that the diffraction pattern can be recorded by a CMOS camera from outside the vacuum chamber.}
\end{figure*}

A cross-sectional view of the TRHEPD experiment with photos of the key components is shown in figure \ref{fig:TRHEPD_overview}. Following the beamline, the beam firstly passes a magnetic field termination, which is described in section \ref{Field Termination}. Inside the subsequent magnetic field-free section, the beam is accelerated by an electrostatic system that consists of 18 electrodes. This system has been designed and optimized to enable measurements with different energies.  A reduced beam energy increases the region of total reflection (see equation \ref{eq:total_reflection}) and might thus be favorable. At NEPOMUC we intend to use an energy of 10 or 15$\thinspace$keV for the remoderated and 10$\thinspace$keV for the twofold remoderated beam.

Inside the Faraday cage, the sample is aligned by a five-axis fine positioning system that can be rotated to change the crystallographic orientation and tilted to enable beam rocking. For an angular step of 0.1$^{\circ}$, we experimentally determined the accuracy to be $\protect\Delta\thinspace\theta$ < 0.02$^{\circ}$ and steps smaller than 0.01$^{\circ}$ can still be resolved. To minimize the influence of the bearing clearance during measurements, we tilt continuously in one direction.

As the intensity of the diffraction pattern is several orders of magnitude lower than in RHEED, the signal must be amplified by a chevron MCP assembly combined with a phosphor screen. We use the commercial MCP \textit{F2226-24P} from \textit{Hamamatsu}, but raised all applied voltages as described in \mbox{section \ref{Detection}}. The diffraction pattern is recorded by a CMOS camera (\textit{Basler acA1920-50gm}) and the images are processed with a Python script.

The vacuum chamber has a base pressure of $\sim\thinspace$2$\cdot$10$\textsuperscript{-9}\thinspace$mbar before bake-out. This pressure can be further reduced by using a homebuilt cryogenic pump, which is basically a cold trap that is filled with liquid nitrogen lasting for over 12$\thinspace$h. After recording a first diffraction pattern, the chamber will be baked out to reach a final pressure in the range of 10$\textsuperscript{-10}\thinspace$mbar.

\subsection{Magnetic Field Termination}
\label{Field Termination}

It is necessary to terminate the magnetic guiding field of the beamline in front of the experiment, because the beam is guided purely electrostatically afterwards. To reduce the influence of stray fields and prevent that positrons follow the field lines, the magnetic field has to end abruptly. Moreover, we do not want to increase the mean transverse energy E$\textsubscript{$\perp$}$ of the beam, worsen its shape or decrease the intensity significantly. As faster positrons are less affected by the field termination, we increase the beam energy from 20$\thinspace$eV up to $\thinspace$2.5$\thinspace$keV before they pass.

Different magnetic field terminations have been applied at NEPOMUC in the past. They range from a simple $\mu$-metal aperture on high voltage to a more sophisticated design that includes thin ferromagnetic stripes in the beam path \citep{Sta08,Pio08,Hug12}. The TRHEPD field termination is inspired by these designs: Within the beam path, we use 25$\thinspace\mu$m thin metallic glass stripes that are long enough to pick up the magnetic flux. They are mechanically clamped in a thin copper frame and orientated to be as parallel as possible to minimize geometrical shadowing (left side figure \ref{fig:TRHEPD_overview}). As this structure leaves gaps of 1$\thinspace$mm, the theoretical transparency of the device is 97.5$\thinspace\%$. Since we plan to bake-out the whole setup at 150$\thinspace^{\circ}$C, we also investigated the temperature behavior of the stripes. A test in an external furnace revealed that the magnetic properties are not affected when baking at 300$\thinspace^{\circ}$C for several hours. This is consistent as the amorphous alloy Fe78-B13-Si9 has a Curie temperature of 415$\thinspace^{\circ}$C. Although, the heat treatment affected the mechanical properties, i.e. the strips became brittle, the clamped structure ensures stability.

The stripes guide the magnetic field lines away from the beam path to be picked up by a $\mu$-metal ring that has a conical structure, as shown in figure \ref{fig:Remoderator}$\thinspace$a) on the right side. Outside the vacuum, we use a cylindrical $\mu$-metal sheet to feed the magnetic field lines back to the beamline and close the circuit. In front of the field termination of the NEPOMUC remoderator, a so-called decompression line is used to gradually reduce the magnetic guiding field \citep{Hug12}. We choose a simpler approach by just decreasing the current of the last flange coil. Experimentally, we observe that this procedure can slightly enhance the beam quality.

\subsection{Optional Transmission-type Remoderator}
\label{Remoderation}

For TRHEPD, the energy spread $\Delta E \approx \Delta E\textsubscript{$\parallel$}$ of the positron beam is of particular importance. Therefore, we define the brightness \textit{B\textsuperscript{+}} as
\begin{equation}
B\textsuperscript{+} = \frac{I}{d\textsuperscript{2} \Theta\textsuperscript{2} E\textsubscript{$\parallel$}\Delta E\textsubscript{$\parallel$}} = \frac{I}{d\textsuperscript{2} E\textsubscript{$\perp$}\Delta E\textsubscript{$\parallel$}},
\label{eq:brightness}
\end{equation}
where \textit{I} is the intensity of the beam, \textit{d} the diameter, $\Theta$ the divergence and \textit{E}\textsubscript{$\parallel$} and \textit{E}\textsubscript{$\perp$} the components of the kinetic energy associated with the momentum parallel and perpendicular to the beam direction.
 
Positron remoderation enables the reduction of the beam diameter, divergence and energy spread at the expense of intensity and can thus increase the brightness of the beam \citep{Mil80}. We expect that the properties of the remoderated NEPOMUC beam are already suitable to obtain a clear diffraction pattern. However, as motivated in \citep{Dod19}, a twofold remoderated beam does not only offer enhanced beam properties but subsequently also leads to a much better illumination of the sample, especially for small angles of incidence. For this reason, the experiment is equipped with an additional remoderator.

\begin{figure}
\includegraphics[trim = 5mm 17mm 5mm 0mm, clip, width=8.1cm]{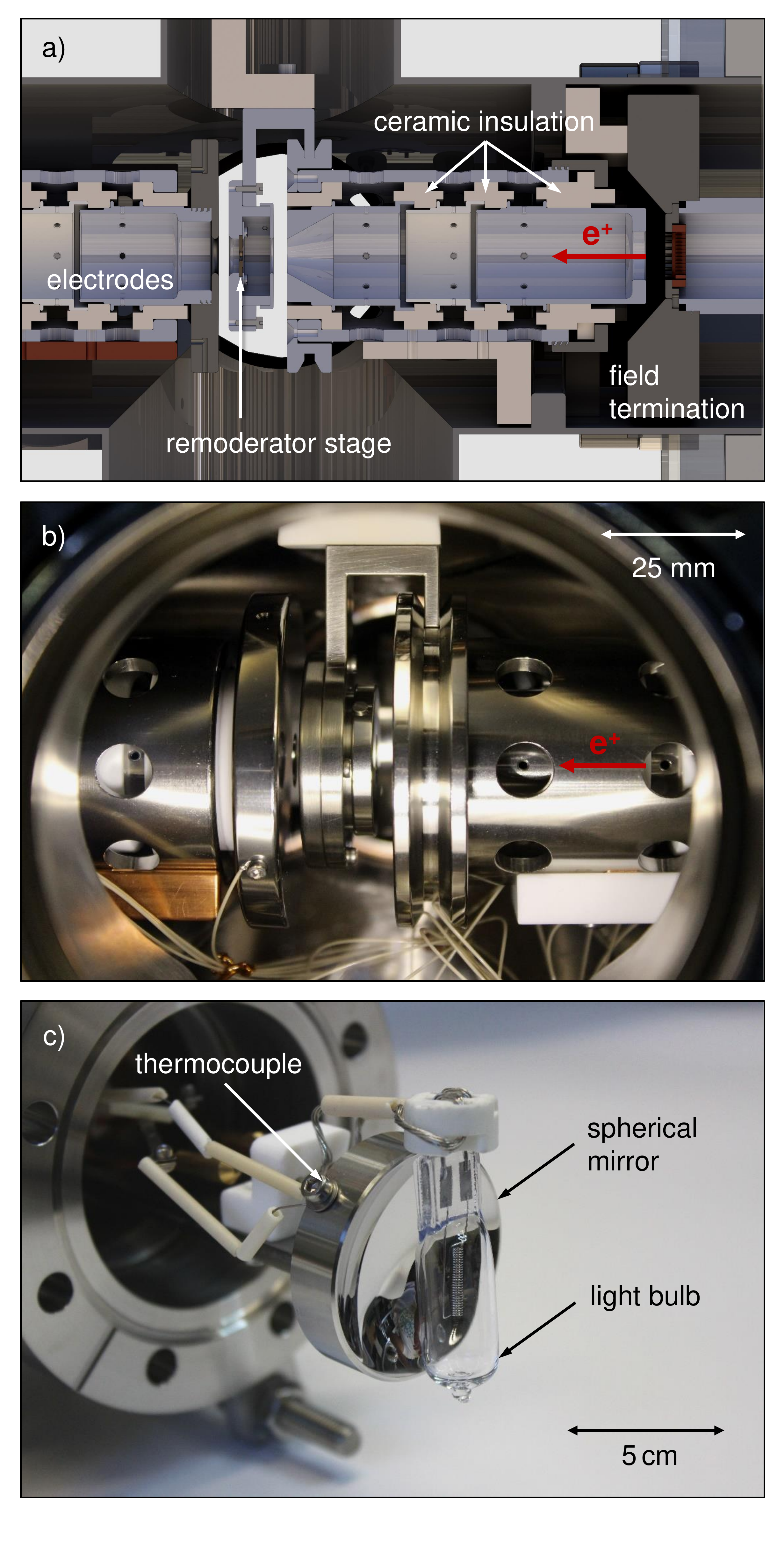}
\setlength\belowcaptionskip{-0.1cm}
\caption{\label{fig:Remoderator} Optional further remoderation using a Ni(100) foil in transmission geometry. The positron beam moves horizontally from right to left. a) Cross-sectional view through the optical axis showing the magnetic field termination, the Einzel lens to adjust the focus and the remoderator stage. b) Photo of the positioned remoderator through the viewport. c) For conditioning in hydrogen atmosphere, the remoderator foil can be heated up to 720$\thinspace^{\circ}$C using a light bulb with spherical mirror.}
\end{figure}

The remoderator stage can be inserted into the beam path from above as depicted in the cross-sectional view in figure \ref{fig:Remoderator}$\thinspace$a). For geometrical reasons and simplicity, the device works in transmission geometry. A spherical fit at the top ensures that the remoderator is positioned concentrically to the electrodes and excludes tilting, so that the surface is perpendicular to the optical axis. A photo of the remoderator stage between the electrostatic lens system is shown in figure \ref{fig:Remoderator}$\thinspace$b). As the distance between remoderator and MCP at the Faraday cage is over 80$\thinspace$cm and the quantum efficiency for $\gamma$-rays is very low, a significant increase of the background due to annihilation in the remoderator foil has not been observed. At the position of the remoderator, it is also possible to insert a MCP from the side to monitor the beam and adjust the focus.

In transmission geometry, mainly tungsten \citep{Osh08} and nickel thin foils \citep{Fuj08,Gig17} have been applied as remoderator materials with efficiencies up to 20$\thinspace\%$. We decided to use a free-standing 100$\thinspace$nm thin, single crystalline Ni(100) foil because nickel can be annealed at much lower temperature and has been shown to yield a narrower energy distribution \citep{Schu86}. To obtain a bright beam, it is crucial to reduce surface roughness and contaminations to a minimum so that the positron work function $\Phi\textsuperscript{+}$ is well defined. In the case of a clean, perfectly aligned remoderator foil with $\Phi _{Ni} ^{+} = -1.4 \thinspace$eV, the only transverse momentum stems from thermal motion, which is roughly 25$\thinspace$meV at room temperature.

The main surface contaminations of nickel are carbon and oxygen, which can be removed by annealing at $\sim\thinspace$1000$\thinspace$K in a hydrogen atmosphere as suggested by Fujinami \emph{et al.} \citep{Fuj08}. Annealing of crystal defects reduces trapping and annihilation in the bulk, but more important, high temperature also induces desorption of surface contaminations. In combination with hydrogen, chemically bound adsorbates can be removed as well \citep{Gig17}. For \emph{in situ} annealing of the remoderator foil, we use a commercial 250$\thinspace$W halogen light bulb with a spherical mirror, as shown in figure \ref{fig:Remoderator}$\thinspace$c). Both are fixed on a rod and can be moved close to the remoderator foil by an edge-welded bellow to maximize the solid angle. The mirror increases the efficiency of the heater and has its focal point well behind the foil to not intensify the thermal gradient along the remoderator. For heating, we use a light bulb instead of a bare filament to prevent tungsten evaporation onto the foil, which would contaminate the remoderator and deteriorate its performance. Since only glass and metal parts are exposed to vacuum, the device is suitable for UHV \citep{Yat15}. The hydrogen flow is controlled with a leak valve and set to reach a pressure in the range of $10^{-4}$ and $10^{-3}\thinspace$mbar.

\subsection{HV-Potentiometer for MCP Potentials}
\label{Detection}

\begin{figure}
\includegraphics[trim = 5mm 3mm 5mm -5mm, clip, width=8.1cm]{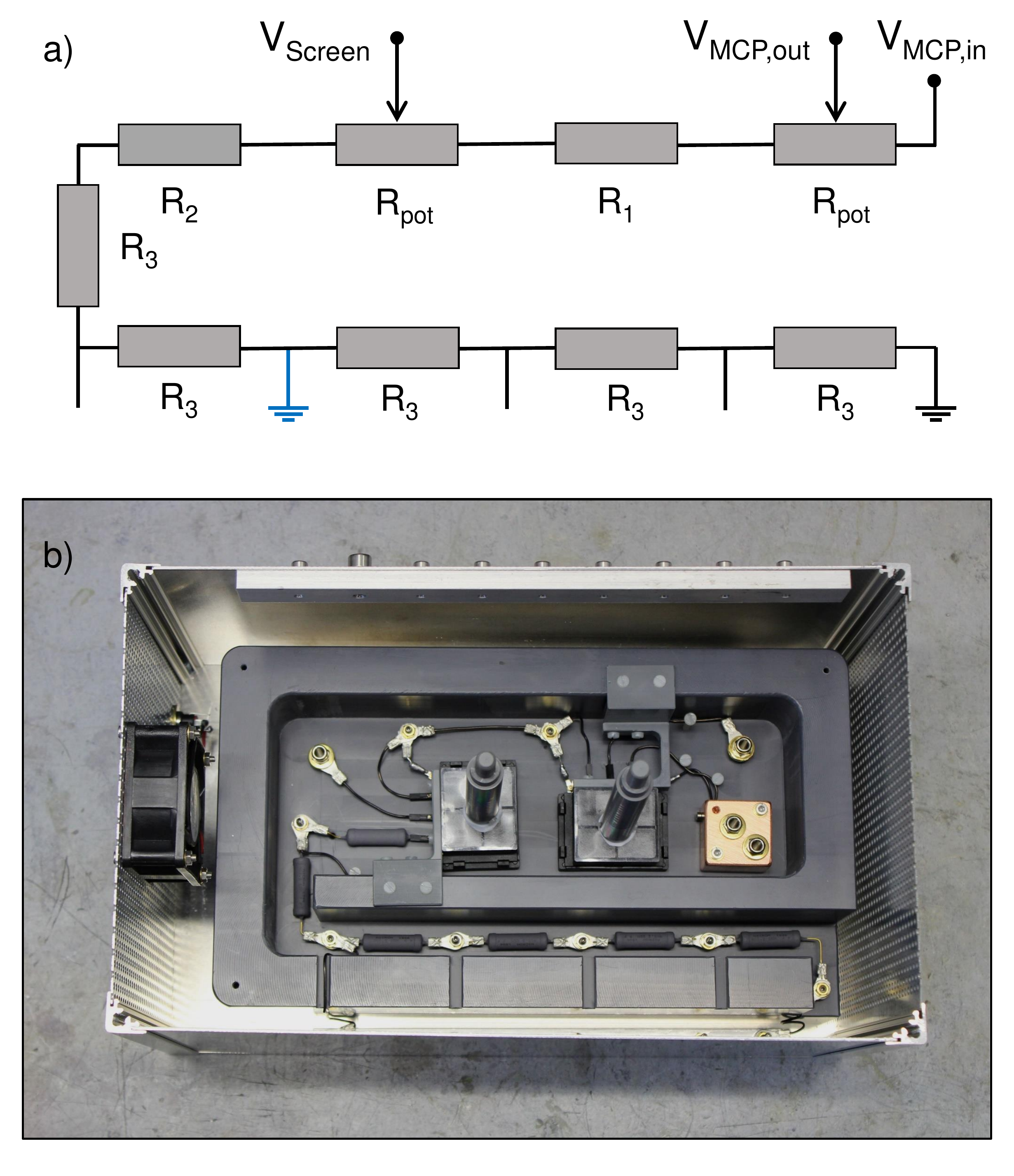}
\setlength\belowcaptionskip{-0.0cm}
\caption{\label{fig:Potentiometer} High voltage potentiometer to provide the MCP potentials, a) block diagram and b) photo. The resistors $R_3=25 \thinspace M\Omega$ can be bypassed at four positions to adjust the potentiometer for different beam energies. The blue bypass in the block diagram corresponds to a beam energy of 15$\thinspace$keV.}
\end{figure}

The MCP assembly is mounted to the Faraday cage to amplify and visualize the diffracted positron beam. The entrance potential of the MCP is defined by the Faraday cage to $\left| V_{MCP, \thinspace in} \right| \thinspace\geq\thinspace$10$\thinspace$kV, depending on the beam energy. As specified in the manual, a maximum potential difference of 2$\thinspace$kV can be applied across the MCP and 4$\thinspace$kV between MCP and phosphor screen. Consequently, we have to raise the potentials $V_{MCP, \thinspace out}$ and $V_{screen}$ accordingly. We decided to provide all three potentials by a single channel of the voltage source using a homemade potentiometer. In the case of a high voltage flashover, the potentials decrease collectively, which reduces the probability to damage the MCP. The potentiometer is shown in figure \ref{fig:Potentiometer}. It consists of several high voltage resistors and two commercial 2.5$\thinspace$kV potentiometers \textit{POC 400} from \textit{Metallux}. These rotary potentiometers are used to adjust the MCP gain and the acceleration towards the screen. They are extended by a PVC rod and can be rotated safely with a grounded knob from outside the housing. To switch for different beam energies, the last resistors of the potentiometer can be bypassed at four positions. In this way, the current through the potentiometer remains the same for all beam energies. The calculated value of 200$\thinspace\mu$A agrees very well with the experimentally observed current of 199.4$\thinspace\mu$A at 15$\thinspace$kV (blue bypass in the block diagram). Such high current is necessary to guarantee stable potentials and compensate possible leak currents and the current that flows across the MCP.

\subsection{Sample Transfer, Heating and Carrier Design}
\label{Samples}

\begin{figure*}
\includegraphics[trim = 0mm 73mm 0mm 0mm, clip, width=17.7cm]{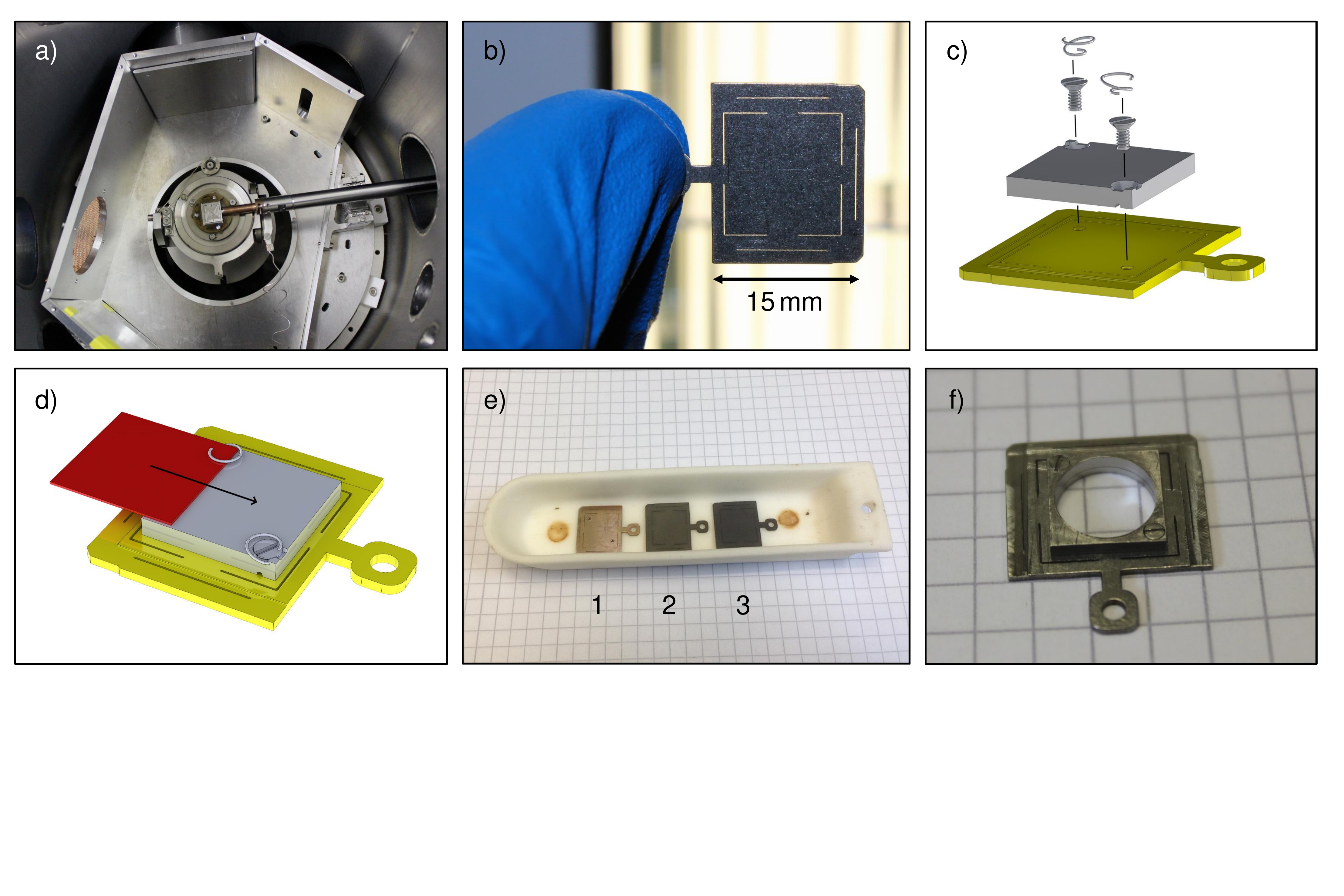}
\setlength\belowcaptionskip{0.0cm}
\caption{\label{fig:Sample Carrier} Transfer and sample carrier design. a) Photo of the open Faraday cage, the transfer shutter can be moved with a push-pull feedthrough from below. A sample garage, which is attached to the shutter at the transfer axis, can accommodate up to four samples. b) To minimize the heat conduction towards the side, the sample carriers have a structure of 150$\thinspace\mu$m thin laser cuts. c) A molybdenum support is screwed to the sample carrier, d) the screw heads hold the clamps that fix the sample. e) In the case of indirect heating, the sample carrier is black oxidized to increase the absorption of the infrared laser light. Different heat treatments in air and oxygen atmosphere have been tested (1-3). f) For direct sample heating we use carriers with a hole in the middle.}
\end{figure*}

At present, sample growth \emph{in situ} is not possible, as the space in the experimental hall of the FRM II is very limited. In the long term, we plan to move the TRHEPD experiment to the neutron guide hall east and connect it with the surface spectrometer SuSpect \citep{May10,Zim14}, which also comprises a system for thin film deposition. For now, samples are grown elsewhere and can be transported with an UHV suitcase that is pumped with a non-evaporable getter (NEG) pump. On a short-timescale of minutes to hours, this allows to maintain an adequate vacuum where, after bake-out, argon is expected to be the main residual gas. In the TRHEPD UHV chamber, the samples can be cleaned using an argon sputter gun and precharacterized with RHEED. As shown in figure \ref{fig:Sample Carrier}$\thinspace$a), the Faraday cage can be opened at the transfer side with a shutter. On the shutter, we mounted a sample garage, where up to four samples can be stored.

After transfer to the UHV chamber, most samples need to be conditioned by a heat treatment to remove adsorbates from the surface. Conventionally, this is done by electron bombardment or resistive heating of the sample or its support. Temperature control during the measurement can also be important, e.g. to induce surface reconstructions. Since the sample stage is on high voltage and difficult to access within the Faraday cage, we use an infrared laser ($\lambda=938\thinspace$nm, 140$\thinspace$W) to heat the sample from below. To reduce the thermal gradient, the laser beam is defocused. The temperature is measured by a pyrometer that has been calibrated with a thermocouple at the sample position.

To minimize the thermal loss due to heat conduction, the sample carrier is made from non-magnetic stainless steel and has a structure of 150$\thinspace\mu$m thin laser cuts, as shown in figure \ref{fig:Sample Carrier}$\thinspace$b). The sample carriers are black oxidized by a heat treatment in an external furnace (10$\thinspace$min at 1000$\thinspace^{\circ}$C in 3$\thinspace$mbar oxygen atmosphere) to increase the absorption of the laser light. A molybdenum support is screwed to the carrier and the sample is clamped by steel springs at two corners as illustrated in figure \ref{fig:Sample Carrier}$\thinspace$c) and d). Indirect heating of the sample leads to a homogeneous temperature distribution, but limits the maximum temperature to roughly 850$\thinspace^{\circ}$C. Samples, that are opaque for the infrared light can also be heated directly, which enables temperatures up to $\sim\thinspace$1000$\thinspace^{\circ}$C. For this, we use a carrier with a hole in the middle, as shown in figure \ref{fig:Sample Carrier}$\thinspace$f).

\section{Positron Beam \textendash{} Simulation and Experiment}
\label{sec:Sim}

Prior to the fabrication of mechanical components, we performed simulations using \textit{COMSOL Multiphysics} to optimize the electrostatic acceleration of the positron beam and the focus on the remoderator foil. Magnetic fields are included in the simulations. To facilitate the calculation, we approximate the system to have rotational symmetry. For this, we neglect the metallic glass strips of the field termination and the extension of the remoderator stage that mechanically ensures the alignment to the optical axis. Tentative simulations for a beam energy of 25$\thinspace$keV have already been published \mbox{in \citep{Dod19}}. The results for reduced energies are now compared with experimental data to characterize the direct beam.

\subsection{Remoderated NEPOMUC Beam}
\label{Remoderated Beam}

The exact properties of the beam at the entrance of the TRHEPD apparatus are unknown. Therefore, we conservatively estimate the beam diameter to be \textit{d}$\thinspace$=$\thinspace$2$\thinspace$mm and the maximum transverse energy to be $E\textsubscript{$\perp$}$= 0.2$\thinspace$eV to simulate the positron trajectories. To account for the increase of the beam diameter due to the gyration in the magnetic guiding field, we set different axial starting positions (in z direction) to obtain a more realistic beam. When the additional remoderator is not applied, the beam has to pass a gap of 20$\thinspace$mm. A simulation of this section is shown in figure \ref{fig:Fokus Remoderator, Sim. und Exp.}$\thinspace$a). The remoderator stage can be moved up by $\sim\thinspace$25$\thinspace$cm and has therefore no effect on the beam. To minimize the influence of the wires and the grounded surrounding, the two electrodes have an outer radius of 34$\thinspace$mm, which is limited by the size of the CF100 crosspiece. This geometry corresponds to a plate capacitor and thus results in a relatively homogeneous field close to the optical axis. The simulation suggests a potential difference of 2.5$\thinspace$kV between the capacitor plates to form a parallel beam. In the subsequent section, we obtain a slightly converging beam by further acceleration. The distance from the last electrode to the sample stage and the MCP at the Faraday cage is 22$\thinspace$cm and 47$\thinspace$cm, respectively. The simulation for \textit{E}$\thinspace$=$\thinspace$15$\thinspace$keV yields, that the beam diameter can be kept below 4$\thinspace$mm over the whole path with a very small divergence of $\Theta\thinspace$<$\thinspace$0.05$\thinspace^{\circ}$. A slightly smaller beam diameter is possible when focusing the beam.

Figure \ref{fig:Fokus Remoderator, Sim. und Exp.}$\thinspace$b) shows the experimentally observed 15$\thinspace$keV direct beam on the MCP together with the intensity distribution along major and minor axis. The beam has an elliptical shape, which is already present at the position of the Ni(100) remoderator and in the beamline downstream the first remoderator. Therefore, the shape is not related to the electrostatic system but rather stems from the non-ideal transport through a magnetic switch in the beamline. Such shape is in general no problem for diffraction as long as the beam is still parallel. However, it is not ideal for rocking curve analysis, which could be improved by adding a suitable aperture. The major and minor axis of the beam diameter (FWHM) were determined to 3.9$\thinspace$mm and 2.7$\thinspace$mm, respectively. Compared to the simulation, the real diameter is smaller than expected which may be due to the rather conservative estimate of the starting conditions or the better transition through the magnetic field termination.

\begin{figure*}
\includegraphics[trim = 0mm 4mm 0mm 0mm, clip, width=17.5cm]{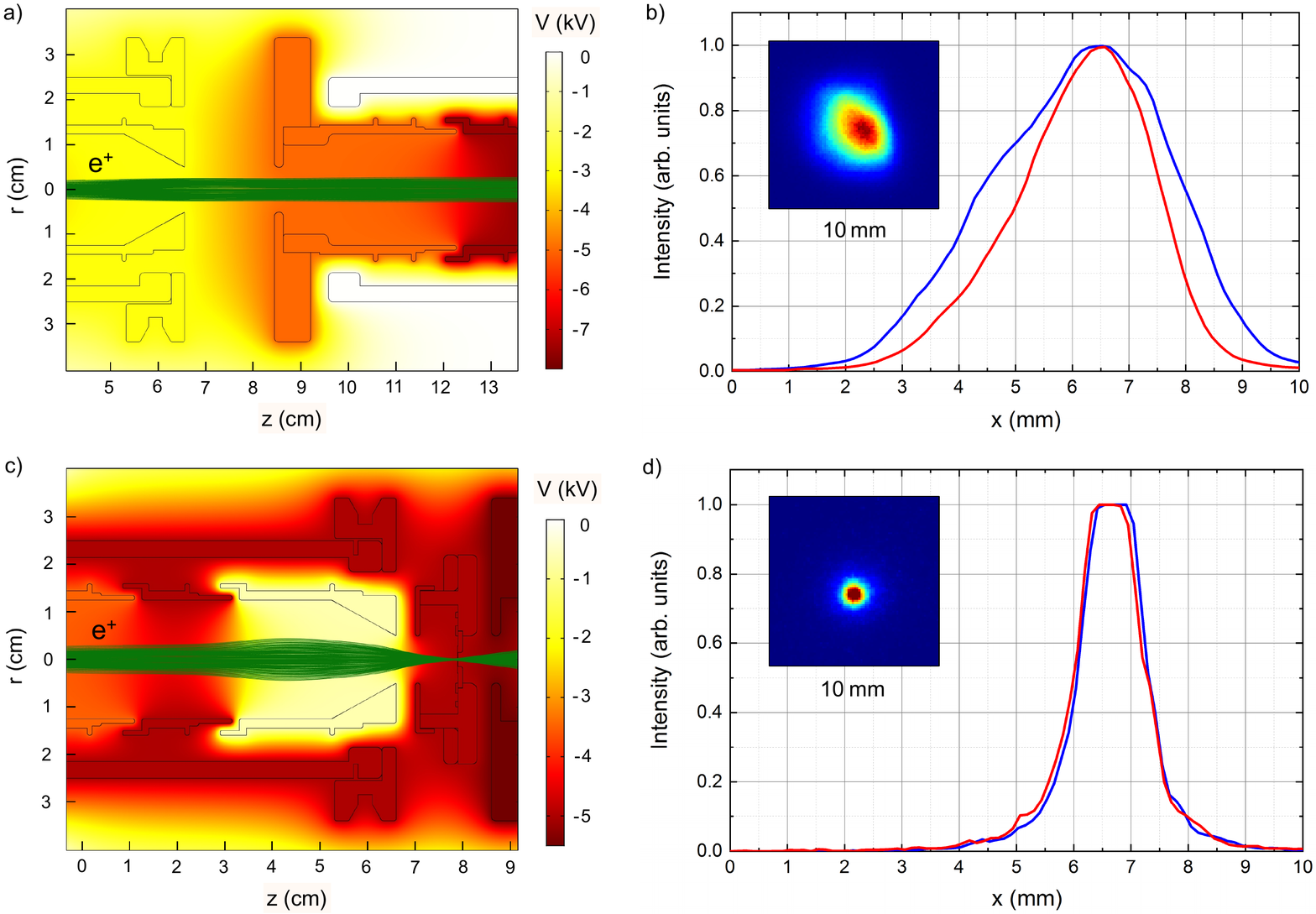}
\setlength\belowcaptionskip{-0.2cm}
\caption{\label{fig:Fokus Remoderator, Sim. und Exp.} Simulation of the positron trajectories in the section of the optional remoderator and experimental characterization of the direct beam. a) Simulation without remoderator stage for \textit{d}$\thinspace$=$\thinspace$2$\thinspace$mm and $E\textsubscript{$\perp$}$= 0.2$\thinspace$eV to obtain a parallel beam. b) Intensity distribution along minor and major axis of the 15$\thinspace$keV remoderated NEPOMUC beam. Insert: a 10$\thinspace$x10mm sized image (linear color scale) of the beam spot depicts the elliptical shape. c) Simulation of the electrostatic focus onto the Ni(100) foil for \textit{d}$\thinspace$=$\thinspace$3$\thinspace$mm and $E\textsubscript{$\perp$}$= 0.2$\thinspace$eV. The simulation predicts a beam diameter of \textit{d} < 0.6$\thinspace$mm at the focal point. d) The intensity distribution of the 10$\thinspace$keV twofold remoderated beam (linear color scale) reveals a circular shape.}
\end{figure*}

\subsection{Twofold Remoderated Beam}
\label{Twofold Remoderated Beam}

To use the twofold remoderated beam, a set of three electrodes accelerates and focuses the beam onto the Ni(100) foil. The positron trajectories have been simulated for different beam diameters \textit{d} (2, 3 and 4$\thinspace$mm) and transverse energies $E\textsubscript{$\perp$}$ (0, 0.2 and 0.5$\thinspace$eV) to also consider particles outside the FWHM of the intensity distribution or with large transverse momentum. Figure \ref{fig:Fokus Remoderator, Sim. und Exp.}$\thinspace$c) shows the simulation for \textit{d}$\thinspace$=$\thinspace$3$\thinspace$mm and \mbox{$E\textsubscript{$\perp$}\thinspace$= 0.2$\thinspace$eV}. Under these conditions, the beam can be focused to a spot with \textit{d} < 0.6$\thinspace$mm. Qualitatively, different starting conditions lead to the same result but can increase the spot size and slightly shift the common focal point. It is possible to adjust the focus onto the remoderator foil by changing the potential of the repulsive electrode. Experimentally, we find the best focus for -$\thinspace$1100$\thinspace$V. The beam is implanted in the Ni foil with an energy of 5$\thinspace$keV, which is a trade-off between a high remoderation yield and a low epithermal positron fraction \citep{Gig17}. To optimize the energy bandwidth of the beam at the expense of a lower intensity, it might be reasonable to reduce the implantation energy to \mbox{3$\thinspace$-$\thinspace$4$\thinspace$keV} at a later stage. After remoderation, the beam properties are well defined and in the subsequent section we simulate the trajectories for \mbox{$E\textsubscript{$\perp$}\thinspace$= 25$\thinspace$meV}. After acceleration to 10$\thinspace$keV, we expect a parallel beam with \textit{d}$\thinspace\sim\thinspace$1.2$\thinspace$mm at the position of the sample and the MCP.\\\\
Figure \ref{fig:Fokus Remoderator, Sim. und Exp.}$\thinspace$d) shows the experimentally observed twofold remoderated beam with the intensity distribution along two perpendicular axes. The shape is almost perfectly circular, as validated by the identical distribution along both axes. The intensity in the middle of the beam spot saturates, so that a precise determination of the beam diameter is not possible. Nevertheless, we can evaluate \textit{d} < 1.3$\thinspace$mm at FWHM, which is very close to the simulation. The beam intensity and remoderation efficiency after conditioning are planed to be determined in the next beamtime.

\section{Conclusion and Outlook}
\label{sec:Conclusion}

After the new TRHEPD setup at NEPOMUC has been successfully assembled and tested, we experimentally investigated the shape and diameter of the direct beam. The results agree well with the preceding simulations, which yield an almost parallel beam. In the next step we will record first diffraction pattern from a relatively stable and less complex surface, e.g. hydrogen terminated Si(111)$\thinspace$-$\thinspace$(1$\times$1). This structure is well understood as it has been investigated by various techniques including TRHEPD \citep{Kaw98,Kaw00}. This will allow us to benchmark the setup and further tune the electrostatic system. Due to the high intensity available at NEPOMUC in combination with the optional twofold remoderation, we expect to observe higher order diffraction spots within comparably short measurement times. After we perform the bake-out of the UHV chamber to reach a base pressure in the range of 10$\textsuperscript{-10}\thinspace$mbar and conduct first quantitative measurements (rocking curve analysis), the setup will be open for user access. In the long-term, it is envisaged to connect the TRHEPD setup with the surface spectrometer SuSpect. This will provide a unique experimental station that allows \emph{in situ} sample growth and enables structural and chemical surface analysis with topmost layer sensitivity using positrons in combination with conventional techniques.

\section*{Acknowledgment}

We thank A. Elovskii, S. Azyzy, K. Wada, I. Mochizuki and \mbox{T. Hyodo} for their support and discussions. The project has been funded by the German BMBF grant no. 05K16WO7, which is gratefully acknowledged.

\section*{Data availability}

Materials and data related to this paper are available from the corresponding author upon reasonable request.


%
%

%



\FloatBarrier

\bibliographystyle{aipnum-cp}
\bibliography{Literaturverzeichnis}

\end{document}